\documentclass[fleqn]{llncs}
\usepackage{amssymb}
\usepackage{sdna}

\title{Network Algebra for Synchronous Dataflow}
\author{J.A. Bergstra\inst{1} \and C.A. Middelburg\inst{1} \and 
        Gh. \c{S}tef\u{a}nescu\inst{2}}
\institute{Informatics Institute, Faculty of Science, \\
           University of Amsterdam,
           Science Park~904, 1098~XH Amsterdam, the Netherlands \\
           \email{J.A.Bergstra@uva.nl,C.A.Middelburg@uva.nl}
           \and
           Department of Computer Science, Faculty of Mathematics and
           Computer Science, \\ 
           University of Bucharest, 
           Strada Academiei 14, Bucharest, Romania \\
           \email{gheorghe.stefanescu@fmi.unibuc.ro}}

\begin{document}
\maketitle

\begin{abstract}
We develop an algebraic theory of synchronous dataflow networks.
First, a basic algebraic theory of networks, called BNA (Basic Network 
Algebra), is introduced.
This theory captures the basic algebraic properties of networks.
For synchronous dataflow networks, it is subsequently extended with 
additional constants for the branching connections that occur between 
the cells of synchronous dataflow networks and axioms for these 
additional constants.
We also give two models of the resulting theory, the one based on stream 
transformers and the other based on processes as considered in process 
algebra.
\begin{classcode}
F.1.1, F.1.2
\end{classcode}
\end{abstract}

\section{Introduction}
\label{introduction}
In this paper we pursue an axiomatic approach to the theory of dataflow
networks.
Network algebra is presented as a general algebraic setting for the
description and analysis of dataflow networks.
A network can be any labelled directed hypergraph that represents some
kind of flow between the components of a system.
For example, flowcharts are networks concerning flow of control and
dataflow networks are networks concerning flow of data.
Assuming that the components have a fixed number of input and output
ports, such networks can be built from their components and (possibly
branching) connections using parallel composition ($\pcomp$),
sequential composition ($ \scomp $) and feedback ($\feed{}$).
The connections needed are at least the identity ($\idn{}$) and
transposition ($\tr{}{}$) connections, but branching connections may
also be needed for specific classes of networks -- e.g.\ the binary
ramification ($\rmfii{}$) and identification ($\idfii{}$) connections
and their nullary counterparts ($\rmfo{}$ and $\idfo{}$) for flowcharts.

An equational theory concerning networks that can be built using the
above-mentioned operations with only the identity and transposition
constants for connections, called BNA (Basic Network Algebra), is
presented.
The axioms of BNA are sound and complete for such networks modulo graph
isomorphism.
BNA is the core of network algebra; for the specific classes of networks
covered, there are additional constants and axioms.
Flowcharts constitute one such class.
BNA is essentially a part of the algebra of flownomials of
C\u{a}z\u{a}nescu and \c{S}tef\u{a}nescu~\cite{CS90} which was developed
for the description and analysis of flowcharts.

In addition to BNA, an extension of BNA for synchronous dataflow
networks is presented.
Process algebra models of BNA and this extension of BNA are given.
These models provide for a very straightforward connection between
network algebra and process algebra.
Unlike process algebra, network algebra is used for describing systems 
as a network of interconnected components.
A clear connection between process algebra and network algebra appears 
to be useful.

For the process algebra models, ACP (Algebra of Communicating Processes)
of Bergstra and Klop~\cite{BK84b} is used, with the silent step and
abstraction, as well as the following additional features: renaming,
conditionals, iteration, prefixing and communication free merge.
Besides, a discrete-time extension of ACP is used to model synchronous 
dataflow networks.

There are strong connections between the work presented in this paper
and other work.
SCAs (Synchronous Concurrent Algorithms), introduced by Thompson and
Tucker in~\cite{TT91}, can be described in the extension of BNA for
synchronous dataflow networks.
In~\cite{BWM94}, Barendregt et al.\ present a model of computable
processes which is essentially a model of BNA; but a slightly different
choice of primitive operations and constants is used.

The paper starts with an outline of network algebra
(Section~\ref{overview}) and some process algebra preliminaries
(Section~\ref{preliminaries}).
Next the signature, the axioms and two models of BNA, including a
process algebra model, are presented (Section~\ref{bna}).
Thereafter the signature, the axioms and two models of the extension of 
BNA for synchronous dataflow networks, including a process algebra 
model, are presented (Section~\ref{na-s}).
Finally, some closing remarks are made (Section~\ref{conclusions}).

The current paper complements~\cite{BMS97a}.
The latter paper is a revision of~\cite{BMS95a} in which the part on 
synchronous dataflow networks has been left out due to space limitations
imposed by the journal.
The current paper is a revision of~\cite{BMS95a} in which the part on 
asynchronous dataflow networks has been left out instead.

\section{Overview of network algebra}
\label{overview}
This section gives an idea of what network algebra is.
The meaning of its operations and constants is explained informally
making use of a graphical representation of networks.
Besides, dataflow networks are presented as a specific class of
networks and the further subdivision into synchronous and asynchronous
dataflow networks is explained in broad outline.
The formal details will be treated in subsequent sections.

\subsection{General}
\label{overview-general}
First the meaning of the operations and constants of BNA
mentioned in Section~\ref{introduction} ($\pcomp$, $\scomp$, $\feed{}$,
$\idn{}$ and $\tr{}{}$) is explained and then the meaning of the 
additional constants for branching connections mentioned in 
Section~\ref{introduction} ($\rmfii{}$, $\rmfo{}$, $\idfii{}$ and 
$\idfo{}$) is explained.

It is convenient to use, in addition to the operations and constants of
BNA, the extensions $\feed{m}$, $\idn{m}$ and $\tr{m}{n}$ of the
feedback operation and the identity and transposition constants.
These extensions are defined by the equations that occur as axioms 
R5--R6, B6 and B8--B9, respectively, of BNA 
(see Section~\ref{bna-sig-axioms}, Table~\ref{tbl-bna}).
They are called the block extensions of the feedback operation and these
constants.
The block extensions of additional constants for branching connections
can be defined in the same vein.

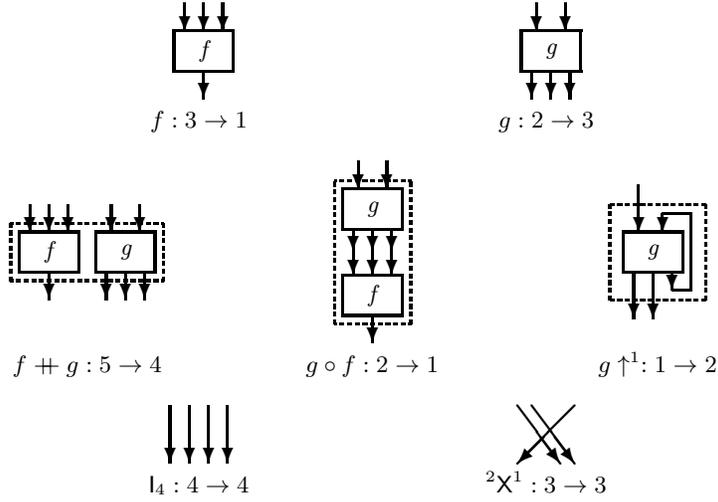
\begin{figure}[tb]
\setlength{\unitlength}{0.005in}
\thicklines
\begin{center}
\begin{minipage}[b]{3.75cm}
\begin{center}
\begin{picture}(70,100)(0,0)
\multiput( 20,100)( 20,  0){3}{\vector( 0,-1){ 30}}
\put( 10, 30){\framebox(60,40){{\em f}}}
\put( 40, 30){\vector( 0,-1){ 30}}
\end{picture}
\\ $f : 3 \to 1$
\end{center}
\end{minipage}
\hspace*{2em}
\begin{minipage}[b]{3.75cm}
\begin{center}
\begin{picture}(70,100)(0,0)
\multiput( 25,100)( 30,  0){2}{\vector( 0,-1){ 30}}
\put( 10, 30){\framebox(60,40){{\em g}}}
\multiput( 20, 30)( 20,  0){3}{\vector( 0,-1){ 30}}
\end{picture}
\\ $g : 2 \to 3$
\end{center}
\end{minipage}
\end{center}

\begin{center}
\begin{minipage}[b]{3.25cm}
\begin{center}
\begin{picture}(160,190)(0,-45)
\multiput( 20,100)( 20,  0){3}{\vector( 0,-1){ 30}}
\put( 10, 30){\framebox(60,40){{\em f}}}
\put( 40, 30){\vector( 0,-1){ 30}}
\multiput(105,100)( 30,  0){2}{\vector( 0,-1){ 30}}
\put( 90, 30){\framebox(60,40){{\em g}}}
\multiput(100, 30)( 20,  0){3}{\vector( 0,-1){ 30}}
\put(  0, 20){\dashbox{4}(160,60){}}
\end{picture}
\\ $f \pcomp g  : 5 \to 4$
\end{center}
\end{minipage}
\hspace*{1em}
\begin{minipage}[b]{3.25cm}
\begin{center}
\begin{picture}(160,190)(-40,0)
\multiput( 25,190)( 30,  0){2}{\vector( 0,-1){ 30}}
\put( 10,120){\framebox(60,40){{\em g}}}
\multiput( 20,120)( 20,  0){3}{\vector( 0,-1){ 25}}
\multiput( 20, 95)( 20,  0){3}{\vector( 0,-1){ 25}}
\put( 10, 30){\framebox(60,40){{\em f}}}
\put( 40, 30){\vector( 0,-1){ 30}}
\put(  0, 20){\dashbox{4}(80,150){}}
\end{picture}
\\ $g \scomp f : 2 \to 1$
\end{center}
\end{minipage}
\hspace*{1em}
\begin{minipage}[b]{3.25cm}
\begin{center}
\begin{picture}(160,190)(-30,-25)
\put( 30,140){\vector( 0,-1){ 50}}
\put( 55,110){\vector( 0,-1){ 20}}
\put( 15, 50){\framebox(60,40){{\em g}}}
\multiput( 25, 50)( 20,  0){2}{\vector( 0,-1){ 50}}
\put( 65, 50){\vector( 0,-1){ 20}}
\put( 65, 30){\line( 1, 0){ 20}}
\put( 85, 30){\line( 0, 1){ 80}}
\put( 85,110){\line(-1, 0){ 30}}
\put(  0, 20){\dashbox{4}(100,100){}}
\end{picture}
\\ $g \feed{1} : 1 \to 2$
\end{center}
\end{minipage}
\end{center}

\begin{center}
\begin{minipage}[b]{3.75cm}
\begin{center}
\begin{picture}(60,60)(0,0)
\multiput(  0, 60)( 20,  0){4}{\vector( 0,-1){ 60}}
\end{picture}
\\ $\idn{4} : 4 \to 4$
\end{center}
\end{minipage}
\hspace*{2em}
\begin{minipage}[b]{3.75cm}
\begin{center}
\begin{picture}(60,60)(0,0)
\multiput(  0, 60)( 15,  0){2}{\vector( 3,-4){ 45}}
\put( 60, 60){\vector(-1,-1){ 60}}
\end{picture}
\\ $\tr{2}{1} : 3 \to 3$
\end{center}
\end{minipage}
\end{center}
\caption{Operations and constants of BNA}
\label{fig-bna}
\end{figure}

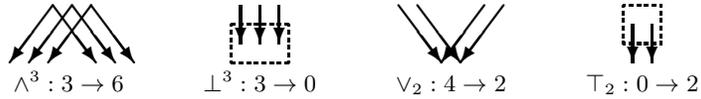
\begin{figure}[tb]
\setlength{\unitlength}{0.005in}
\thicklines
\begin{center}
\begin{minipage}[b]{2.00cm}
\begin{center}
\begin{picture}(110,60)(0,0)
\multiput( 30,  0)( 20,  0){3}{
\begin{picture}(40,60)(0,0)
\put(  0, 60){\vector(-3,-4){ 45}}
\put(  0, 60){\vector( 3,-4){ 45}}
\end{picture}
}
\end{picture}
\\ $\rmfii{3} : 3 \to 6$
\end{center}
\end{minipage}
\hspace*{1em}
\begin{minipage}[b]{2.00cm}
\begin{center}
\begin{picture}(110,60)(-25,0)
\multiput( 10, 60)( 20,  0){3}{\vector( 0,-1){ 40}}
\put(  0,  0){\dashbox{4}(60,40){}}
\end{picture}
\\ $\rmfo{3} : 3 \to 0$
\end{center}
\end{minipage}
\hspace*{1em}
\begin{minipage}[b]{2.00cm}
\begin{center}
\begin{picture}(110,60)(0,0)
\put(  0, 60){\vector( 3,-4){ 45}}
\put( 90, 60){\vector(-3,-4){ 45}}
\put( 20, 60){\vector( 3,-4){ 45}}
\put(110, 60){\vector(-3,-4){ 45}}
\end{picture}
\\ $\idfii{2} : 4 \to 2$
\end{center}
\end{minipage}
\hspace*{1em}
\begin{minipage}[b]{2.00cm}
\begin{center}
\begin{picture}(110,60)(-35,0)
\put(  0, 20){\dashbox{4}(40,40){}}
\multiput( 10, 40)( 20,  0){2}{\vector( 0,-1){ 40}}
\end{picture}
\\ $\idfo{2} : 0 \to 2$
\end{center}
\end{minipage}
\end{center}
\caption{Additional constants for branching connections}
\label{fig-na}
\end{figure}

In Figure~\ref{fig-bna}, the meaning of the operations and constants of
BNA (including the block extensions) is illustrated by means of a
graphical representation of networks.
We write $f : k \to l$ to indicate that network $f$ has $k$ input ports
and $l$ output ports; $k \to l$ is called the sort of $f$.
The input ports are numbered $1,\ldots,k$ and the output ports
$1,\ldots,l$.
In the graphical representation, they are considered to be numbered
from left to right.
The networks are drawn with the flow moving from top to bottom.
Note that the symbols for the feedback operation and the constants fit
with this graphical representation.
In Figure~\ref{fig-na}, the meaning of (block extensions of) the
additional constants for branching connections mentioned in
Section~\ref{introduction} is illustrated by means of a graphical 
representation.
The symbols for these additional constants fit with the graphical
representation as well.

\begin{figure}[tb]
\setlength{\unitlength}{0.005in}
\begin{center}
\begin{picture}(360,280)(0,0)
\thicklines
\multiput( 5,60)( 0,80){3}{\vector(1, 0){40}}
\multiput(40, 0)(80, 0){4}{
\begin{picture}(80,280)(0,0)
\multiput( 0,40)( 0,80){3}{\framebox(40,40){{\em f}}}
\multiput(40,60)( 0,80){3}{\vector(1, 0){40}}
\multiput(20,40)( 0,80){4}{\vector(0,-1){40}}
\put(20,140){\oval(80,280)[br]}
\put(20,140){\oval(80,280)[tr]}
\end{picture}
}
\end{picture}
\end{center}
\caption{A regular network}
\label{fig-network}
\end{figure}
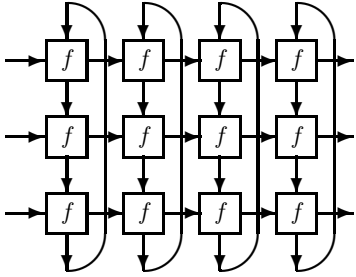

The operations and constants illustrated above allow to represent all
networks (cf.~\cite{Ste86}).
For example,
\begin{center}
$
\begin{array}{l}
r_{k,l} =
(({\scomp}^{k-1}_{i=1}(\idn{k-i} \pcomp {\pcomp}^{i}\!f \pcomp \idn{l-i})
   \scomp {} \\ \phantom{r_{k,l} = ((}
  {\scomp}^{l-k}_{i=0}(\idn{i} \pcomp {\pcomp}^{k}\!f \pcomp \idn{l-k-i})
   \scomp {} \\ \phantom{r_{k,l} = ((}
  {\scomp}^{1}_{i=k-1}(\idn{l-i} \pcomp {\pcomp}^{i}\!f \pcomp \idn{k-i})
   \scomp \tr{l}{k}) \feed{l}\;,
\end{array}
$
\end{center}
where $k<l$ and $f : 2 \to 2$, represent a regular network
(some abbreviations are used here: iterated sequential composition
$ \scomp^n_{i=m} f_i = f_m \scomp \ldots \scomp f_n$ and parallel
composition to the $n$th $\pcomp^{n} f = f \pcomp \dots \pcomp f$
($n$ times)).
The instance $r_{3,4}$ is illustrated in Figure~\ref{fig-network}.

The graphical illustration of the meaning of the operations and
constants of BNA in Figure~\ref{fig-bna} gives intuitive grounds for
the soundness of the axioms of BNA (see Section~\ref{bna-sig-axioms},
Table~\ref{tbl-bna}) for the intended network model.
Similarly, the illustration of the meaning of the additional constants
for branching connections in Figure~\ref{fig-na} makes most additional
axioms for these constants (see Section~\ref{bna-sig-axioms},
Table~\ref{tbl-na}) plausible.

\subsection{Dataflow networks}
In the case of dataflow networks, the components are also called cells.
The identity connections are called wires and the transposition
connections are viewed as crossing wires.
The cells are interpreted as processes that consume data at their input
ports, compute new data, deliver the new data at their output ports,
and then start over again.
The wires are interpreted as queues of some kind.
The classical kinds considered are firstly queues that deliver data with
a neglectible delay and never contain more than one datum, and secondly
unbounded, delaying queues.
In this paper, they are called {\em minimal stream delayers\/} and
{\em stream delayers}, respectively.
A stream is a sequence of data consumed or produced by a component of a
dataflow network.
A flow (of data) is a transformation of a tuple of streams into a tuple
of streams.
A wire behaves as an identity flow.
If the wire is a stream delayer, data pass through it with a time delay.
If the wire is a minimal stream delayer, data enter and leave it with a
neglectible delay -- i.e.\ within the same time slice in case time is
divided into time slices with the length of the time unit used.

In synchronous dataflow networks, the wires are minimal stream delayers.
Basic to synchronous dataflow is that there is a global clock.
On ticks of the clock, cells can start up the consumption of exactly
one datum from each of their input ports and the production of exactly
one datum at each of their output ports.
A cell that started up with that completes the production of data before
the next tick, and it completes the consumption of data as soon as a
new datum has been delivered at all input ports.
On the first tick following the completion of both, the cell concerned
starts up again.
In order to start the synchronous dataflow network, every cell has, for
each of its output ports, an initial datum available to deliver on the
initial tick.
The underlying idea of synchronous dataflow is that computation takes a
good deal of time, whereas storage and transport of data takes a
neglectible deal of time.
Phrased differently, data always pass through a wire between two
consecutive ticks of the global clock.
So minimal stream delayers fit in exactly with this kind of dataflow
networks.
The semantics of synchronous dataflow networks turns out to be rather
simple and unproblematic.

In asynchronous dataflow networks, the wires are stream delayers.
The underlying idea of asynchronous dataflow is that computation as
well as storage and transport of data takes a good deal of time, which
is sometimes more realistic for large systems.
In such cases, it is favourable to have computation driven by the
arrival of the data needed -- instead of by clock ticks.
Therefore, there is no global clock in an asynchronous dataflow network.
Cells may independently consume data from their input ports,
compute new data, and deliver the new data at their output ports.
Because it means that there may be data produced by cells but not yet
consumed by other cells, this needs wires that are able to buffer an
arbitrary amount of data.
So stream delayers fit in exactly with this kind of dataflow networks.
However, the semantics of asynchronous dataflow networks turns out to
be rather problematic.
The main semantic problem is a time anomaly, known as the
Brock-Ackermann anomaly.
With feedback, timing differences in producing data may become
important and the time anomaly actually shows that delaying queues do
not perfectly fit in with that.
Besides, the unbounded queues needed to keep an arbitrary amount of data
are unrealistic.
Note that a synchronous dataflow network can be viewed as a extreme
case of an asynchronous one, where the queues never contain more than
one datum.

Dataflow networks also need branching connections.
Their branching structure is more complex than the branching structure
of flowcharts.
In case of flowcharts, there is a flow of control which is always at
one point in the flowchart concerned.
In consequence, the interpretation of the branching connections is
rather obvious.
However, in case of dataflow networks, there is a flow of data which is
everywhere in the network.
Hence, the interpretation of the branching connections is not
immediately clear.
In this paper, two kinds of interpretation are considered.
For the binary branching connections, they are the
{\em copy}/{\em equality~test\/} interpretation and the
{\em split}/{\em merge\/} interpretation.
The first kind of interpretation fits in with the idea of permanent
flows of data which naturally go in all directions at branchings.
Synchronous dataflow reflects this idea most closely.
The second kind of interpretation fits in with the idea of intermittent
flows of data which go in one direction at branchings.
Asynchronous dataflow reflects this idea better.
In order to distinguish between the branching constants with these
different interpretations, different symbols for $\rmfii{m}$ and
$\idfii{m}$ are used: $\cp{m}$ and $\eq{m}$ for the copy/equality~test
interpretation, $\split{m}$ and $\merge{m}$ for the split/merge
interpretation.
Likewise, different symbols for the nullary counterparts $\rmfo{m}$ and
$\idfo{m}$ are used: $\ssink{m}$ and $\ssour{m}$ versus $\asink{m}$ and
$\asour{m}$.
$\ssink{m}$ and $\asink{m}$ are called {\em sink\/} and
{\em dummy sink}, respectively; and $\ssour{m}$ and $\asour{m}$ are
called {\em source} and {\em dummy source}, respectively.

%

In the synchronous case, with minimal stream delayers as identity
connections and the copy/equality~test interpretation of the branching
connections, it turns out that two axioms for $\rmfii{m}$ and 
$\idfii{m}$ are not valid.
Fortunately the others together with two new axioms give a complete set 
of axioms.
The asynchronous case is somewhat problematic owing to the time anomaly
that occurs in the model outlined above.
The asynchronous case is treated separately in~\cite{BMS97a}.

Dataflow networks have been extensively studied, see
e.g.~\cite{BWM94,Boh84,BA81,Bro88,Jon89,Kah74,Kok87,Rus89}.

\section{Process algebra preliminaries}
\label{preliminaries}
This section gives a brief summary of the ingredients of process
algebra which make up the basis for the process algebra models
presented in Sections~\ref{bna} and \ref{na-s}.
We will suppose that the reader is familiar with them.
Appropriate references to the literature are included.

We will make use of \acpt, which is an extension of ACP~\cite{BK84b} 
with abstraction based on branching bisimulation~\cite{GW96}.
In \acpt, processes can be composed from actions, the inactive process 
($\dead$) and the silent step ($\tau$) by sequential composition 
($\seqc$), alternative composition ($\altc$), parallel composition 
($\parc$), encapsulation ($\encap{H}$), and abstraction ($\abstr{I}$).
For a systematic introduction to \acpt, the reader is referred
to~\cite{BW90}.
We will use the following abbreviation.
Let $(P\sb{i})\sb{i \in I}$ be a indexed set of process expressions
where $I = \set{i\sb1,\ldots,i\sb{n}}$.
Then, we write $\sum_{i \in I} P\sb{i}$ for
$P\sb{i\sb1} \altc \ldots \altc P\sb{i\sb{n}}$ if $n > 0$ and $\dead$
if $n = 0$.

We will also use some of the features added to ACP in~\cite{BB94a}:
\begin{description}
\item[\it Renaming]
We will use the renaming operator $\rnm{f}$.
Here $f$ is a function that renames actions into actions, $\dead$ or
$\tau$.
The expression $\rnm{f}(P)$ denotes the process $P$ with every occurrence
of an action $a$ replaced by $f(a)$.
So the most crucial equation from the axioms for the renaming operator is 
$\rnm{f}(a) = f(a)$.
\item[\it Conditionals]
We will use the two-armed conditional operator $\cond{\ \ }$.
The expression $P \cond{\,b\,} Q$, is to be read as
$\kw{if}\, b\, \kw{then}\, P\, \kw{else}\, Q$.
The most important equations derivable from the axioms for the two-armed 
conditional operator are $X \cond{\True} Y = X$ and 
$X \cond{\False} Y = Y$.
\item[\it Early input prefixing]
We will use the early input action prefixing operators \sloppy
($er_i(v) \pref {}$) and their generalization to a process prefixing 
operator ($\pref$).
The most important equation derivable from the axioms for the early 
input action prefixing operators is 
$er_i(v) \pref X = \sum_{d \in D} r_i(d) \seqc X[d/v]$ (it is assumed 
that a fixed but arbitrary finite set $D$ of data has been given).
\item[\it Process prefixing]
We will use the process prefixing operator mainly to express parallel input:
$(er_1(v_1) \parc \dots \parc er_n(v_n)) \pref P$.
We have:
\begin{center}
\footnotesize
$
\begin{array}[t]{lll}
(er_1(v_1) \parc er_2(v_2)) \pref P
 & = & \Altc{d_1 \in D} r_1(d_1) \seqc (er_2(v_2) \pref P[d_1/v_1]) \\
 & + & \Altc{d_2 \in D} r_2(d_2) \seqc (er_1(v_1) \pref P[d_2/v_2])\;,
\vsp \\
(er_1(v_1) \parc er_2(v_2) \parc er_3(v_3)) \pref P
 & = & \Altc{d_1 \in D} r_1(d_1)\seqc((er_2(v_2) \parc
                           er_3(v_3)) \pref P[d_1/v_1]) \\
 & + & \Altc{d_2 \in D} r_2(d_2)\seqc((er_1(v_1) \parc
                           er_3(v_3)) \pref P[d_2/v_2]) \\
 & + & \Altc{d_3 \in D} r_3(d_3)\seqc((er_1(v_1) \parc
                           er_2(v_2)) \pref P[d_3/v_3])\;,
\vsp \\
\mbox{etc.}
\end{array}
$
\end{center}
\item[\it Communication free merge]
We will use the communication free merge operator ($\cfm$). 
This operator is in fact one of the synchronisation merge operators 
$\parc_H$ of CSP, which are also added to ACP in~\cite{BB94a}, viz.\  
$\parc_\emptyset$.
Communication free merge can also be expressed in terms of parallel
composition, encapsulation and renaming.
The most crucial equations from the axioms for the communication free 
merge operator are $P \cfm Q = P \cflm Q \altc Q \cflm P$ and
$a \seqc P \cflm Q = a \seqc (P \cfm Q)$.
\end{description}

Moreover, we will make use of \acpdrtt, which is an extension of 
\acpdrt\ with abstraction based on branching bisimulation.
\acpdrt\ in turn is an extension of ACP with discrete relative timing.
In \acpdrtt, time is considered to be divided into slices indexed by 
natural numbers.
These time slices represent time intervals of a length which corresponds
to the time unit used.
In \acpdrtt, we have the additional constants $\csl{a}$ (for each action 
$a$), $\csl{\tau}$ and $\csl{\dead}$, and the delay operator $\delay$.
The process $\asl{a}$ is $a$ performed in any time slice and $\csl{a}$
is $a$ performed in the current time slice.
Similarly, $\csl{\tau}$ is a silent step performed in the current time
slice and $\csl{\dead}$ is inaction in the current time slice.
The expression $\delay(P)$ denotes the process $P$ delayed one time 
slice.
The process $\asl{a}$ is recursively defined by the equation
$X = \csl{a} \altc \delay(X)$.
In a parallel composition $P\sb1 \parc \dots \parc P\sb{n}$ the
transition to the next time slice is a simultaneous transition of 
$P\sb1, \dots, P\sb{n}$.
For example, $\csl{\dead} \parc \delay(\csl{b})$ will never perform
$\csl{b}$ because $\csl{\dead}$ can neither be delayed nor performed, so
$\csl{\dead} \parc \delay(\csl{b}) = \csl{\dead}$.
However, $\csl{a} \parc \delay(\csl{b}) = \csl{a} \seqc \delay(\csl{b})$.
For a systematic introduction to \acpdrtt, the reader is referred
to~\cite{BM02a}.

We will also use the above-mentioned features in the setting of \acpdrtt.
The integration of renaming, conditionals, and communication free merge 
in the discrete time setting is obvious.
The integration of early input prefixing and process prefixing may seem 
less clear at first sight, but the relevant equations are simply
$\csl{er}_i(v) \pref X = \sum_{d \in D} \csl{r}_i(d) \seqc X[d/v]$ and
$\delay(X) \pref Y = \delay(X \pref Y)$.

\section{Basic network algebra}
\label{bna}
BNA is essentially the part of the algebra of flownomials~\cite{CS90}
that is common to various classes of networks.
In particular, it is common to flowcharts and dataflow networks.
The additional constants, needed for branching connections, differ
however from one class to another.
In this section, BNA is presented.
First of all, the signature and axioms of BNA are given.
The extension of BNA to the algebra of flownomials is also addressed
here.
In addition, two models of BNA are described: a data transformer model
and a process algebra model.
In subsequent sections, an extension of BNA for synchronous dataflow 
networks is provided.

\subsection{Signature and axioms of BNA}
\label{bna-sig-axioms}
\subsubsection*{Signature}
In network algebra, networks are built from other networks -- starting
with atomic components and a variety of connections.
Every network $f$ has a sort $k \to l$, where $k,l \in \Nat$, associated
with it.
To indicate this,  we use the notation $f : k \to l$.
The intended meaning of the sort $k \to l$ is the set of networks with
$k$ input ports and $l$ output ports.
So $f : k \to l$ expresses that $f$ has $k$ input ports and $l$ output
ports.

The sorts of the networks to which an operation of network algebra is
applied determine the sort of the resulting network.
In addition, there are restrictions on the sorts of the networks to
which an operation can be applied.
For example, sequential composition can not be applied to two networks
of arbitrary sorts because the number of output ports of one should
agree with the number of input ports of the other.

The signature of BNA is as follows:
\begin{center}
\footnotesize
\begin{tabular}{l@{\quad}c@{\quad}l}
 Name  & Symbol & Arity \\[-1.25ex]
\multicolumn{3}{l}{\rule{.99\textwidth}{.125mm}} \svsp \\
{\bf Operations:} \svsp \\
  parallel composition   & $\pcomp$ &
  $(k \to l) \x (m \to n) \to (k+m \to l+n)$ \\
  sequential composition & $\scomp$  &
  $(k \to l) \x (l \to m) \to (k \to m)$     \\
  feedback               & $\feed{}$ &
  $(m+1 \to n+1) \to (m \to n)$                 \svsp \\
{\bf Constants:} \svsp \\
  identity      & $\idn{}$  & $1 \to 1$         \\
  transposition & $\tr{}{}$ & $2 \to 2$         \\
\multicolumn{3}{l}{\rule{.99\textwidth}{.125mm}} \\
\end{tabular}
\end{center}
Here $k,l,m,n$ range over $\Nat$.
This means, for example, that there is an instance of the sequential
composition operator for each $k,l,m \in \Nat$.

As mentioned in Section~\ref{overview}, we will also use the block
extensions of feedback, identity and transposition.
The arity of these auxiliary operations and constants is as follows:
\begin{center}
\footnotesize
\begin{tabular}{c@{\quad}l}
Symbol & Arity \\[-1.25ex]
\multicolumn{2}{l}{\rule{.99\textwidth}{.125mm}} \svsp \\
$\feed{l}$  & $(m+l \to n+l) \to (m \to n)$ \\
$\idn{m}$   & $m \to m$                     \\
$\tr{m}{n}$ & $m+n \to n+m$                 \\
\multicolumn{2}{l}{\rule{.99\textwidth}{.125mm}} \\
\end{tabular}
\end{center}

\subsubsection*{Axioms}
The axioms of BNA are given in Table~\ref{tbl-bna}.
\begin{table}[tb]
\caption{Axioms of BNA}
\label{tbl-bna}
\rule{.99\textwidth}{.125mm}
\begin{center}
\footnotesize
\begin{tabular}{l@{\quad}ll}
  B1 & $f \pcomp (g \pcomp h) = (f \pcomp g) \pcomp h$ \\
  B2 & $\idn{0} \pcomp f = f = f \pcomp \idn{0}$ \\
  B3 & $f \scomp (g \scomp h) = (f \scomp g) \scomp h$  \\
  B4 & $\idn{k} \scomp f = f = f \scomp \idn{l}$ \\
  B5 & $(f \pcomp f') \scomp (g \pcomp g')
        = (f \scomp g) \pcomp (f' \scomp g')$ \\
  B6 & $\idn{k} \pcomp \idn{l} = \idn{k+l}$ \\
  B7 & $\tr{k}{l} \scomp \tr{l}{k} = \idn{k+l}$ \\
  B8 & $\tr{k}{0} = \idn{k}$ \\
  B9 & $\tr{k}{l+m}
        = (\tr{k}{l} \pcomp \idn{m}) \scomp
          (\idn{l} \pcomp \tr{k}{m})$ \\
 B10 & $(f \pcomp g) \scomp \tr{m}{n} = \tr{k}{l} \scomp (g \pcomp f)$
     & \ \ \  for $f:k \to m,\ g:l \to n$ \svsp \\
  R1 & $g \scomp (f\feed{m})
        = ((g \pcomp \idn{m}) \scomp f) \feed{m}$ \\
  R2 & $(f\feed{m}) \scomp g
        = (f \scomp (g \pcomp \idn{m} )) \feed{m}$ \\
  R3 & $f \pcomp (g \feed{m}) = (f \pcomp g) \feed{m}$  \\
  R4 & $(f \scomp (\idn{l} \pcomp g)) \feed{m}
        = ((\idn{k} \pcomp g) \scomp f) \feed{n}$ 
     & \ \ \  for $f : k+m \to l+n,\ g : n \to m$ \\
  R5 & $f\feed{0} = f$ \\
  R6 & $(f\feed{l}) \feed{k} = f \feed{k+l}$ \svsp \\
  F1 & $\idn{k} \feed{k} = \idn{0}$ \\
  F2 & $\tr{k}{k} \feed{k} = \idn{k}$ \\
\end{tabular}
\end{center}
\rule{.99\textwidth}{.125mm}
\end{table}
The axioms B1--B6 for $\pcomp$, $\scomp$ and $\idn{m}$ define a
strict monoidal category; and together with the additional axioms B7--B10
for $\tr{m}{n}$, they define a {\em symmetric\/} strict monoidal
category (ssmc for short).
The remaining axioms R1--R6 and F1--F2 characterize $\feed{l}$.
The axioms R5--R6, B6 and B8--B9 can be regarded as the defining
equations of the block extensions of $\feed{}$, $\idn{}$ and $\tr{}{}$,
respectively.

The axioms of BNA are sound and complete for networks modulo graph
isomorphism (cf.~\cite{Ste86}).
Using the graphical representation of Section~\ref{overview-general},
it is easy to see that the axioms in Table \ref{tbl-bna} are sound.
By means of the axioms of BNA, each expression can be brought into a
normal form
$$
((\idn{m} \pcomp x_1 \pcomp \dots \pcomp x_k) \scomp f)
                                        \feed{m_1 + \dots + m_k}\;,
$$
where the $x_i : m_i \to n_i$ ($i \in [k]$)%
\footnote{We write $[n]$, where $n \in \Nat$, for $\set{1,\ldots,n}$.}
are the atomic components of the network and
$f : m + n_1 + \dots + n_k \to n + m_1 + \dots + m_k$ is a bijective
connection.
A network is uniquely represented by a normal form expression up to a
permutation of $x_1,\ldots,x_k$.
The completeness of the axioms of BNA now follows from the fact that
these permutations in a normal form expression are deducible from the
axioms of BNA as well.

As a first step towards the stream transformer and process algebra
models for synchronous dataflow networks described in 
Section~\ref{na-s}, a data transformer model and a process algebra 
model of BNA are provided immediately after the connection with the 
algebra of flownomials has been addressed.

\subsubsection*{Extension to the algebra of flownomials}
The algebra of flownomials is essentially\footnote{
For naming ports, an arbitrary monoid is used in the algebra of
flownomials whereas the monoid of natural numbers is used in BNA.}
a conservative extension of BNA.
Recall that the algebra of flownomials was not developed for dataflow
networks, but for flowcharts.
The signature of the algebra of flownomials is obtained by extending the
signature of BNA as follows with additional constants for branching
connections:
\begin{center}
\footnotesize
\begin{tabular}{l@{\quad}c@{\quad}l@{\quad}l}
 Name  & Symbol & Arity & Instances \\[-1.25ex]
\multicolumn{4}{l}{\rule{.99\textwidth}{.125mm}} \svsp \\
{\bf Additional constants:} \svsp \\
  ramification  & $\rmf{}{k}$ & $1 \to k$ &
  $\left\{
   \begin{array}{lll}
   \rmfii{} & := & \rmf{}{2} \\
   \rmfo{}  & := & \rmf{}{0}
   \end{array}
   \right.$                           \svsp \\
  identification & $\idf{k}{}$ & $k \to 1$ &
  $\left\{
   \begin{array}{lll}
   \idfii{} & := & \idf{2}{} \\
   \idfo{}  & := & \idf{0}{}
   \end{array}
   \right.$                            \\
\multicolumn{4}{l}{\rule{.99\textwidth}{.125mm}} \\
\end{tabular}
\end{center}
We will restrict our attention to the instances for $k = 0$ and
$k = 2$, i.e. $\rmfii{}$, $\rmfo{}$, $\idfii{}$ and $\idfo{}$.
The other instances can be defined in terms of them:
$$
\begin{array}[t]{lll}
\rmf{}{k+1} & = & \rmfii{} \scomp (\rmf{}{k} \pcomp \idn{})\;, \svsp \\
\idf{k+1}{} & = & (\idf{k}{} \pcomp \idn{}) \scomp \idfii{}\;.
\end{array}
$$
It follows from these definitions, together with the axioms A3 and A7
of the algebra of flownomials (see Table~\ref{tbl-na}), that
$\rmf{}{1} = \idf{1}{} = \idn{}$.

We will use the block extensions of $\rmfii{}$, $\rmfo{}$, $\idfii{}$
and $\idfo{}$.
The arity of these auxiliary constants is as follows:
\begin{center}
\footnotesize
\begin{tabular}{c@{\quad}l}
 Symbol & Arity \\[-1.25ex]
\multicolumn{2}{l}{\rule{.99\textwidth}{.125mm}} \svsp \\
$\rmfii{m}$ & $m \to 2m$  \\
$\rmfo{m}$  & $m \to 0$   \\
$\idfii{m}$ & $2m \to m$  \\
$\idfo{m}$  & $0 \to m$   \\
\multicolumn{2}{l}{\rule{.99\textwidth}{.125mm}} \\
\end{tabular}
\end{center}
The axioms for the additional constants of the algebra of flownomials
are given in Table~\ref{tbl-na}.
\begin{table}[tb]
\caption{Additional axioms for flowcharts}
\label{tbl-na}
\rule{.99\textwidth}{.125mm}
\begin{center}
\footnotesize
\begin{tabular}{l@{\quad}l}
  A1 & $(\idfii{m} \pcomp \idn{m}) \scomp \idfii{m}
        = (\idn{m} \pcomp \idfii{m}) \scomp \idfii{m}$ \\
  A2 & $\tr{m}{m} \scomp \idfii{m} = \idfii{m}$ \\
  A3 & $(\idfo{m} \pcomp \idn{m}) \scomp \idfii{m} = \idn{m}$ \\
  A4 & $\idfii{m} \scomp \rmfo{m} = \rmfo{m} \pcomp \rmfo{m}$ \\[1.5ex]
  A5 & $\rmfii{m} \scomp (\rmfii{m} \pcomp \idn{m})
        = \rmfii{m} \scomp (\idn{m} \pcomp \rmfii{m})$ \\
  A6 & $\rmfii{m} \scomp \tr{m}{m} = \rmfii{m}$ \\
  A7 & $\rmfii{m} \scomp (\rmfo{m} \pcomp \idn{m}) = \idn{m}$ \\
  A8 & $\idfo{m} \scomp \rmfii{m} = \idfo{m} \pcomp \idfo{m}$ \\[1.5ex]
  A9 & $\idfo{m} \scomp \rmfo{m} = \idn{0}$ \\
  A10 & $\idfii{m} \scomp \rmfii{m}
               = (\rmfii{m} \pcomp \rmfii{m}) \scomp
                 (\idn{m} \pcomp \tr{m}{m} \pcomp \idn{m}) \scomp
                 (\idfii{m} \pcomp \idfii{m})$ \\
  A11 & $\rmfii{m} \scomp \idfii{m} = \idn{m}$ \\[1.5ex]
  A12 & $\idfo{0} = \idn{0}$ \\
  A13 & $\idfo{m+n} = \idfo{m} \pcomp \idfo{n}$ \\
  A14 & $\idfii{0} = \idn{0}$ \\
  A15 & $\idfii{m+n}
         = (\idn{m} \pcomp \tr{n}{m} \pcomp \idn{n}) \scomp
           (\idfii{m} \pcomp \idfii{n})$ \\[1.5ex]
  A16 & $\rmfo{0} = \idn{0}$ \\
  A17 & $\rmfo{m+n} = \rmfo{m} \pcomp \rmfo{n}$ \\
  A18 & $\rmfii{0} = \idn{0}$ \\
  A19 & $\rmfii{m+n}
         = (\rmfii{m} \pcomp \rmfii{n}) \scomp
           (\idn{m} \pcomp \tr{m}{n} \pcomp  \idn{n})$ \\[1.5ex]
  F3 & $\idfii{m}\feed{m} = \rmfo{m}$ \\
  F4 & $\rmfii{m}\feed{m} = \idfo{m}$ \\
  F5 & $((\idn{m} \pcomp \rmfii{m}) \scomp
                (\tr{m}{m} \pcomp \idn{m}) \scomp
                (\idn{m} \pcomp \idfii{m})) \feed{m} = \idn{m}$ \\
\end{tabular}
\end{center}
\rule{.99\textwidth}{.125mm}
\end{table}
These axioms where chosen in order to describe the branching structure
of flowcharts.
The axioms A12--A19 can be regarded as the defining equations of the
block extentions of $\rmfii{}$, $\rmfo{}$, $\idfii{}$ and $\idfo{}$.

The standard model for the interpretation of flowcharts is the model
$\Rel(D)$ of relations over a set $D$ (cf.~\cite{CS90,Ste87b}).
All axioms of the algebra of flownomials (Tables~\ref{tbl-bna}
and~\ref{tbl-na}) hold in this model.
The algebraic structure defined by the axioms of BNA (Table~\ref{tbl-bna})
was introduced in~\cite{Ste86} under the name of {\em biflow}.
In~\cite{Ste94} it is called {\em $a\alpha$-ssmc with feedback}.
The algebraic structure defined by the axioms of the algebra of
flownomials (Tables~\ref{tbl-bna} and~\ref{tbl-na}) is called
{\em $d\delta$-ssmc with feedback\/} in~\cite{Ste94}.

\subsection{Data transformer model of BNA}
\label{bna-rel}
In this subsection, a data transformer model of BNA is described.
A parallel data transformer $f : m \to n$ acts on an $m$-tuple of input
data and produces an $n$-tuple of output data.
Parallel composition, sequential composition and feedback operators as
well as identity and transposition constants are defined on parallel
data transformers.
All axioms of BNA (Table~\ref{tbl-bna}) hold in the resulting model.

\bdfn (data transformer model of BNA)
\label{dfn-rel}

\noindent
A {\em parallel data transforming relation} $f \in \rel(S)(m,n)$
is a relation
$$
f \subseteq S^m \x  S^n\;,
$$
where $S$ is a set of data.
$\rel(S)$ denotes the indexed family of data transforming relations
$(\rel(S)(m,n))_{\textstyle \Nat \x \Nat}$.

The operations and constants of BNA are defined on $\rel(S)$ as
follows:
\begin{center}
\footnotesize
\begin{tabular}{@{}lll@{}}
\multicolumn{3}{l}{Notation} \\[-1.25ex]
\multicolumn{3}{l}{\rule{.99\textwidth}{.125mm}} \svsp \\
 $f \pcomp g$ & $\in \rel(S)(m+p,n+q)$
            & for $f \in \rel(S)(m,n)$, $g \in \rel(S)(p,q)$ \\
 $f \scomp g$      & $\in \rel(S)(m,p)$
            & for $f \in \rel(S)(m,n)$, $g \in \rel(S)(n,p)$ \\
 $f \feed{p}$      & $\in \rel(S)(m,n)$
            & for $f \in \rel(S)(m+p,n+p)$ \vsp \\
 $\idn{n}$    & $\in \rel(S)(n,n)$              \\
 $\tr{m}{n}$  & $\in {\sf \rel }(S)(m+n,n+m)$   \\
\multicolumn{3}{l}{\rule{.99\textwidth}{.125mm}}
\end{tabular}
\svsp \\
$
\begin{array}{lll}
\multicolumn{3}{l}{\mbox{Definition}\footnotemark} \\[-1.25ex]
\multicolumn{3}{l}{\rule{.99\textwidth}{.125mm}} \svsp \\
f \pcomp g    & = &
\set{\pair{x \cat y}{z \cat w} \where
     x \in S^m \And y \in S^p \And z \in S^n \And w \in S^q \And
     \pair{x}{z} \in f \And \pair{y}{w} \in g} \\
f \scomp g & = &
\set{\pair{x}{y} \where
     x \in S^m \And y \in S^p \And
     \exists z \in S^n \st \pair{x}{z} \in f \And \pair{z}{y} \in g} \\
f \feed{p} & = &
\set{\pair{x}{y} \where
     x \in S^m \And y \in S^n \And
     \exists z \in S^p \st \pair{x \cat z}{y \cat z} \in f} \vsp \\
\idn{n}    & = & 
\set{\pair{x}{x} \where x \in S^n} \\
\tr{m}{n}  & = & 
\set{\pair{x \cat y}{y \cat x} \where x\in S^m \And y\in S^n} \\
\multicolumn{3}{l}{\rule{.99\textwidth}{.125mm}}
\end{array}
$
\end{center}
\footnotetext
{Let $x = \tup{x_1,\ldots,x_m}$ and $y = \tup{y_1,\ldots,y_n}$ be
 tuples.
 Then we write $x \cat y$ for the tuple
 $\tup{x_1,\ldots,x_m,y_1,\ldots,y_n}$.
 Moreover, we often write $\pair{x_1}{x_2}$ instead of $\tup{x_1,x_2}$.}
\edfn

The definitions of the operations and constants of BNA given above are 
very straightforward.
Note that the data transformer model defined here has a global crash 
property: if a component of a network fails to produce output, the whole 
network fails to produce output.

\bthm
\label{thm-rel}
$(\rel(S),\pcomp,\scomp,\feed{},\idn{},\tr{}{})$ is a model of BNA.
\ethm
\bproof
The proof is a matter of straightforward calculation using only
elementary set theory.
\eproof

Additional branching constants can be defined such that the resulting
expanded model satisfies most axioms of the algebra of flownomials
(Tables~\ref{tbl-bna} and~\ref{tbl-na}).
One such set of branching constants is closely related to the one that
is used in the design of (nondeterministic) SCAs~\cite{TT91}.
The corresponding expanded model is principally the stream transformer
model for synchronous dataflow networks described in Section~\ref{na-s}
where an abstraction is made from the internals of the transformers:
arbitrary data is transformed instead of streams of data.
However, $\idfo{m}$ must be interpreted as $\ssour{m}$ in this data
transformer model, to keep up relationships with SCAs, whereas it is
interpreted as $\asour{m}$ in the stream transformer model for
synchronous dataflow networks.

\subsection{Process algebra model of BNA}
\label{bna-proc}
Network algebra can be regarded as being built on top of process
algebra.

Let $D$ be a fixed, but arbitrary, finite set of data.
$D$ is a parameter of the model.
The processes use the standard actions $r_i(d)$, $s_i(d)$ and
$c_i(d)$ for $d \in D$ only.
They stand for read, send and communicate, respectively, datum $d$ at
port $i$.
On these actions, communication is defined such that
$r_i(d) \commm s_i(d) = c_i(d)$\linebreak[2] (for all $i \in \Nat$ and $d \in D$).
In all other cases, it yields $\dead$.

We write $H(i)$, where $i \in \Nat$, for the set
$\set{r_i(d) \where d \in D} \union \set{s_i(d) \where d \in D}$
and $I(i)$ for $\set{c_i(d) \where d \in D}$.
In addition, we write
$H(i,j)$ for $H(i) \union H(j)$,
$H(i + [k])$ for $H(i + 1) \union \dots \union H(i + k)$ and
$H(i + [k],j + [l])$ for $H(i + [k]) \union H(j + [l])$.
The abbreviations $I(i,j)$, $I(i + [k])$ and $I(i + [k],j + [l])$ are
used analogously.

$in(i/j)$ denotes the renaming function defined by
$$
\begin{array}[t]{llll}
in(i/j)(r_i(d))    & = & r_j(d) & \mbox{for } d \in D\;, \\
in(i/j)(a)         & = & a      &
                  \mbox{for } a \notin \set{r_i(d) \where d \in D}\;.
\end{array}
$$
So $in(i/j)$ renames port $i$ into $j$ in read actions.
$out(i/j)$ is defined analogously, but renames send actions.
We write $in(i + [k]/j + [k])$ for
$in(i + 1/j + 1) \comp \ldots \comp in(i + k/j + k)$ and
$in([k]/j + [k])$ for $in(0 + [k]/j + [k])$.
The abbreviations $out(i + [k]/j + [k])$ and $out([k]/j + [k])$ are
used analogously.
\bdfn (process algebra model of BNA)
\label{dfn-proc}

\noindent
A {\em network\/} $f \in \proc(D)(m,n)$ is a triple
$$
f = (m,n,P)\;,
$$
where $P$ is a process with actions in
$\set{r_i(d) \where i \in [m] \And d \in D} \union
 \set{s_i(d) \where i \in [n] \And\linebreak[2] d \in D}$.
$\proc(D)$ denotes the indexed family of sets
$(\proc(D)(m,n))_{\textstyle \Nat \x \Nat}$.

A {\em wire\/} is a network $\idn{} = (1,1,w^1_1)$, where $w^1_1$
satisfies for all networks $f = (m,n,P)$ and $u,v > \max(m,n)$: 
\begin{center}
\begin{tabular}{lll}
(P1) &
$\abstr{I(u,v)}(\encap{H(v,u)}(w^u_v \parc w^v_u)) \cfm P = P\;,$ \vsp \\
(P2) &
$\abstr{I(u,v)}
 (\encap{H(u,v)}((\rnm{in(i/u)}(P)  \cfm w^i_v) \parc w^v_u)) = P$
     & for all $i \in [m]\;,$ \vsp \\
(P3) &
$\abstr{I(u,v)}
 (\encap{H(u,v)}((\rnm{out(j/v)}(P) \cfm w^u_j) \parc w^v_u)) = P$
     & for all $j \in [n]\;,$ 
\end{tabular}
\end{center}
where $w^u_v = \rnm{in(1/u)}(\rnm{out(1/v)}(w^1_1))$.
\pagebreak[2]

%

The operations and constants of BNA are defined on $\proc(D)$ as follows:
\begin{center}
\small
\begin{tabular}{lll}
\multicolumn{3}{l}{Notation} \\[-1.25ex]
\multicolumn{3}{l}{\rule{.99\textwidth}{.125mm}} \svsp \\
 $f \pcomp g$ & $\in \proc(D)(m+p,n+q)$
            & for $f \in \proc(D)(m,n)$, $g \in \proc(D)(p,q)$ \\
 $f \scomp g$ & $\in \proc(D)(m,p)$
            & for $f \in \proc(D)(m,n)$, $g \in \proc(D)(n,p)$ \\
 $f \feed{p}$      & $\in \proc(D)(m,n)$
            & for $f \in \proc(D)(m+p,n+p)$                    \vsp \\
 $\idn{n}$         & $\in \proc(D)(n,n)$ \\
 $\tr{m}{n}$       & $\in \proc(D)(m+n,n+m)$ \\
\multicolumn{3}{l}{\rule{.99\textwidth}{.125mm}}
\end{tabular}
\svsp \\
$
\begin{array}{lll}
\multicolumn{3}{l}{\mbox{Definition}} \\[-1.25ex]
\multicolumn{3}{l}{\rule{.99\textwidth}{.125mm}} \svsp \\
(m,n,P) \pcomp (p,q,Q)  & = & (m+p,n+q,R)\;, \\
\multicolumn{3}{l}{
\mbox{where } R = P \cfm \rnm{in([p]/m+[p])}(\rnm{out([q]/n+[q])}(Q))}
\vsp \\
(m,n,P) \scomp (n,p,Q)  & = & 
(m,p,\abstr{I(u+[n],v+[n])}(\encap{H(u+[n],v+[n])}(R)))\;, \\ 
\multicolumn{3}{l}{\mbox{where } u = \max(m,p), v = u + n, \mbox{ and}} \\
\multicolumn{3}{l}{
R = (\rnm{out([n]/u+[n])}(P) \cfm \rnm{in([n]/v+[n])}(Q)) \parc
    w^{u+1}_{v+1} \parc \dots \parc  w^{u+n}_{v+n}} 
\vsp \\
(m+p,n+p,P) \feed{p}    & = & 
(m,n,\abstr{I(u+[p],v+[p])}(\encap{H(u+[p],v+[p])}(R)))\;, \\
\multicolumn{3}{l}{\mbox{where } u = \max(m,n), v = u + p, \mbox{ and}} \\
\multicolumn{3}{l}{
R = \rnm{in(m+[p]/v+[p])}(\rnm{out(n+[p]/u+[p])}(P)) \parc
    w^{u+1}_{v+1} \parc \dots \parc  w^{u+p}_{v+p}}
\end{array}
$
\lvsp \\
$
\begin{array}{l@{}l@{}ll@{}l@{}ll}
\idn{n} & \Eq & 
(n,n,w^1_1 \cfm \dots \cfm w^n_n)                      & \mbox{if } n > 0 \\
& & 
(0,0,\abstr{I(1,2)}(\encap{H(1,2)}(w^1_2 \parc w^2_1))) & \mbox{otherwise}
\vsp \\
\tr{m}{n} & \Eq & 
(m+n,n+m,
 w^1_{n+1} \cfm \dots \cfm w^m_{n+m} \cfm
 w^{m+1}_1 \cfm \dots \cfm w^{m+n}_n)              & \mbox{if } m + n > 0 \\
& &
(0,0,\abstr{I(1,2)}(\encap{H(1,2)}(w^1_2 \parc w^2_1))) & \mbox{otherwise}
\\
\multicolumn{7}{l}{\rule{.99\textwidth}{.125mm}}
\end{array}
$
\end{center}
\edfn

The conditions (P1)--(P3) on wires given above are rather obscure at 
first sight, but they are equivalent to the axioms B2 and B4 of BNA:
(P1) corresponds to $\idn{0} \pcomp f = f = f \pcomp \idn{0}$,
(P2) to $\idn{m} \scomp f = f$, and (P3) to $f = f \scomp \idn{n}$.
The definitions of sequential composition and feedback illustrate
clearly the differences between the mechanisms for using ports in
network algebra and process algebra.
In network algebra the ports that become internal after composition are
hidden.
In process algebra based models these ports are still visible; a special
operator must be used to hide them.
For typical wires, $\abstr{I(1,2)}(\encap{H(1,2)}(w^1_2 \parc w^2_1))$
equals $\dead$, $\tau \seqc \dead$ or $\csl{\tau} \seqc \dead$ (the
latter only in case \acpdrtt\ is used).

In the description of a process algebra model of BNA given above, all
constants and operators used are common to \acpt\ and \acpdrtt\ or
belong to a few of their mutual (conservative) extensions mentioned in
Section~\ref{preliminaries} (viz.\ renaming and communication free
merge).
As a result, we can specialize this general model for a specific kind
of networks using either \acpt\ or \acpdrtt; with further extensions at
need.
On the other hand, we can obtain general results on these process
algebra models: results that only depend on properties that are common
to \acpt\ and \acpdrtt\ or properties of the mutual extensions used
above.
\bthm
\label{thm-proc}
$(\proc(D),\pcomp,\scomp,\feed{},\idn{},\tr{}{})$ is a model of BNA.
\ethm
\bproof
According to~\cite{Ste94}, there is an algebra equivalent to BNA (the
algebra of LR-flow over \Bi), but having two renumbering operations,
for (bijectively) renumbering input ports and output ports, instead of
the transposition constant and the sequential composition operation of
BNA.
Renumbering is just renaming in the corresponding process algebra model.
The crucial axioms concerning the constant $\idn{n}$ in the equational
theory of that algebra follow immediately from the conditions (P1)--(P3)
on wires in Definition~\ref{dfn-proc}.
For quite a few axioms from this equational theory, the proof that they
are satisfied by the process algebra model is a matter of simple
calculation using only elementary properties of renaming, communication
free merge, or parallel composition and renaming.
For the remaining axioms, reminiscent of the axioms R1--R4 of BNA, the
proof is a matter of straightforward calculation using in addition
properties of parallel composition and encapsulation or abstraction.
All properties concerned are common to \acpt\ and \acpdrtt\ or
properties of the mutual extensions used in Definition~\ref{dfn-proc}.
\eproof

If we select a specific wire, such as $\msd$ in Section~\ref{na-s}, 
we have obtained a model of BNA if the conditions (P1)--(P3) are 
satisfied by the wire concerned.

\section{Synchronous dataflow networks}
\label{na-s}
In this section, an extension of BNA for synchronous dataflow networks
is presented.
First of all, the additional constants and axioms for synchronous
dataflow are given.
Next, the adaptation of the data transformer model of
Section~\ref{bna-rel} to synchronous dataflow networks, resulting in a
stream transformer model for synchronous dataflow, is described.
Finally, the specialization of the process algebra model of
Section~\ref{bna-proc} for synchronous dataflow networks is described.

\subsection{Additional constants and axioms}
\label{na-s-sig-axioms}
The signature of the extension of BNA for synchronous dataflow networks
is obtained by extending the signature of BNA as follows with additional
constants for branching connections:
\begin{center}
\footnotesize
\begin{tabular}{l@{\quad}c@{\quad}l}
 Name  & Symbol & Arity \\[-1.25ex]
\multicolumn{3}{l}{\rule{.99\textwidth}{.125mm}} \svsp \\
{\bf Additional constants:} \svsp \\
  copy          & $\cp{m}$    & $m  \to 2m$  \\
  sink          & $\ssink{m}$ & $m  \to  0$  \\
  equality test & $\eq{m}$    & $2m \to  m$  \\
  dummy source  & $\asour{m}$ & $0  \to  m$  \\
\multicolumn{3}{l}{\rule{.99\textwidth}{.125mm}} \\
\end{tabular}
\end{center}
The symbols $\cp{m}$, $\ssink{m}$ and $\eq{m}$ indicate that the
copy/equality~test interpretation is intended here.
For technical reasons, which are explained at the end of
Section~\ref{na-s-qrel}, $\asour{m}$ is used instead of $\ssour{m}$.

The axioms for these additional constants are given in Table~\ref{tbl-na-s}.
\begin{table}[tb]
\caption[t]{Additional axioms for synchronous dataflow networks}
\label{tbl-na-s}
\rule{.99\textwidth}{.25mm}
\begin{center}
\footnotesize
\begin{tabular}{l@{\quad}l}
  A1 & $(\eq{m} \pcomp \idn{m}) \scomp \eq{m}
        = (\idn{m} \pcomp \eq{m}) \scomp \eq{m}$ \\
  A2 & $\tr{m}{m} \scomp \eq{m} = \eq{m}$ \\
  A3$^\circ$
     & $(\asour{m} \pcomp \idn{m}) \scomp \eq{m}
        = \ssink{m} \scomp \asour{m}$ \\
  A4 & $\eq{m} \scomp \ssink{m} = \ssink{m} \pcomp \ssink{m}$ \svsp \\
  A5 & $\cp{m} \scomp (\cp{m} \pcomp \idn{m})
        = \cp{m} \scomp (\idn{m} \pcomp \cp{m})$ \\
  A6 & $\cp{m} \scomp \tr{m}{m} = \cp{m}$ \\
  A7 & $\cp{m} \scomp (\ssink{m} \pcomp \idn{m}) = \idn{m}$ \\
  A8 & $\asour{m} \scomp \cp{m} = \asour{m} \pcomp \asour{m}$ \svsp \\
  A9 & $\asour{m} \scomp \ssink{m} = \idn{0}$ \\
  A10 & $\eq{m} \scomp \cp{m}
               = (\cp{m} \pcomp \cp{m}) \scomp
                 (\idn{m} \pcomp \tr{m}{m} \pcomp \idn{m}) \scomp
                 (\eq{m} \pcomp \eq{m})$ \\
  A11 & $\cp{m} \scomp \eq{m} = \idn{m}$ \svsp \\
  A12 & $\asour{0} = \idn{0}$ \\
  A13 & $\asour{m+n} = \asour{m} \pcomp \asour{n}$ \\
  A14 & $\eq{0} = \idn{0}$ \\
  A15 & $\eq{m+n}
         = (\idn{m} \pcomp \tr{n}{m} \pcomp \idn{n}) \scomp
           (\eq{m} \pcomp \eq{n})$ \svsp \\
  A16 & $\ssink{0} = \idn{0}$ \\
  A17 & $\ssink{m+n} = \ssink{m} \pcomp \ssink{n}$ \\
  A18 & $\cp{0} = \idn{0}$ \\
  A19 & $\cp{m+n}
         = (\cp{m} \pcomp \cp{n}) \scomp
           (\idn{m} \pcomp \tr{m}{n} \pcomp \idn{n})$ \svsp \\
  F3 & $\eq{m} \feed{m} = \ssink{m}$ \\
  F4 & $\cp{m}\feed{m}  = \asour{m}$ \\
  F5$^\circ$ &
       $((\idn{m} \pcomp \cp{m}) \scomp
         (\tr{m}{m} \pcomp \idn{m}) \scomp
         (\idn{m} \pcomp \eq{m})) \feed{m}
        = \ssink{m} \scomp \asour{m}$ \\
\end{tabular}
\end{center}
\rule{.99\textwidth}{.25mm}
\end{table}
These axioms agree with those for the additional constants of the
algebra of flownomials (Table~\ref{tbl-na}) with two exceptions: A3 and
F5 are replaced by A3$^\circ$ and F5$^\circ$.

In the next two subsections, the models introduced in Section~\ref{bna}
are specialized to describe the semantics of the synchronous dataflow
networks.

\subsection{Stream transformer model for synchronous dataflow}
\label{na-s-qrel}
In this subsection, an adaptation of the data transformer model of BNA
(Section~\ref{bna-rel}) for synchronous dataflow is given.

In Section~\ref{bna-rel}, no assumptions about the nature of the
transformers were made.
Here the nature of the transformers needed for synchronous dataflow
networks is made precise, resulting in the definition of quasiproper
stream transformers.
The feedback operation is adapted to reflect a special characteristic of
feedback in synchronous dataflow networks: data in the feedback loop
produced in one time slice is not used to produce new data before the
next time slice.

The model $\rel(S)$ of Section~\ref{bna-rel} is a general model.
In case of dataflow, streams of data are transformed.
This means that
$$
S = \iseqof{(D \union \set{\tick})} = \Nat \to (D \union \set{\tick})
$$
for some set of data $D$, $\tick \notin D$.
For a stream $x \in S$ and $k \in \Nat$, $x(0..k)$ is the initial 
segment of $x$ of length $k + 1$ and $x(k)$ is the datum occurring in 
$x$ on the $k$-th tick of the global clock if $x(k) \in D$.
The absence of a datum is represented by $\tick$; so $x(k) = \tick$
indicates that no datum occurs in stream $x$ on the $k$-th tick.
This may happen, for example, with the equality test $\eq{1}$: no datum
is delivered on the $k$-th tick unless equal data are offered at its
input ports on that tick.
Owing to this approach to deal with the absence of data, it is quite
natural in case of synchronous dataflow to look at finite streams as
infinite ones where no datum occurs from a certain tick.
This point of view has the additional advantage that the relevant
definitions can be kept simple.
However, it is unnatural to uphold this view-point for asynchronous
dataflow.

The stream transformers used to model the cells in synchronous dataflow
networks have a ``dependency on the past'' property which is captured by
the following definition.
\bdfn (proper stream transformer)
\label{dfn-proper}

\noindent
A stream transformer $f \in \rel(S)(m,n)$ is {\em proper\/}
(or {\em determined by the past\/}) if
\begin{center}
$
\begin{array}[t]{l}
\forall x  \in S^m  \st  \forall x' \in S^m  \st {} \\ \quad
  \set{y(0) \where y \in S^n, \pair{x}{y} \in f}
  = \set{  y'(0) \where y' \in S^n, \pair{x'}{y'} \in f} \And {} \\ \quad\,
 \forall k  \in \Nat \st x(0..k) = x'(0..k) \hsp \Implies {} \\ \qquad\,
  \set{y(0..k+1) \where y \in S^n, \pair{x}{y} \in f}
  = \set{  y'(0..k+1) \where y' \in S^n, \pair{x'}{y'} \in f}\;.
\end{array}
$
\end{center}
\edfn

Note that this property reduces at the beginning to a ``constant output
initially'' property.

The proper stream transformers fail to include constants for connections
such as $\idn{}$, $\tr{}{}$, $\cp{}$ and $\eq{}$, because their intended
meaning is to let data pass through them with a neglectible delay.
Because at least the constants $\idn{}$ and $\tr{}{}$ are necessary in
order to define a network algebra, stream transformers built from proper
stream transformers and stream transformers with input and output ports
that are directly connected must be allowed.
The resulting stream transformers are called quasiproper stream
transformers.
A similar notion is used in~\cite{BWM94}.
\bdfn (direct connection)
\label{dfn-direct}

\noindent
Two ports $i \in [m]$ and $j \in [n]$ are {\em directly connected\/}
via a stream transformer $f \in {\sf \rel }(S)(m,n)$ if
\begin{center}
$
\begin{array}[t]{l}
\forall (x_1,\dots,x_m) \in S^m \st
\forall (y_1,\dots,y_n) \in S^n \st {} \\ \hsp
 \pair{(x_1,\dots,x_m)}{(y_1,\dots,y_n)} \in f \hsp \Implies \hsp
 x_i = y_j\;.
\end{array}
$
\end{center}
We write $dc(f)$ for the set
$\set{(i,j) \where
      i \mbox{ is directly connected with } j \mbox{ via } f}$.

A stream transformer $f \in \rel(S)(m,n)$ is a {\em direct connection\/}
if
$$
\forall i \in [m] \st \exists j \in [n] \st (i,j) \in dc(f) \And
\forall j \in [n] \st \exists i \in [m] \st (i,j) \in dc(f)\;.
$$
\edfn
\bdfn (quasiproper stream transformer)
\label{dfn-quasiproper}

\noindent
A stream transformer in $\rel(S)(m,n)$ is {\em quasiproper\/}
if it can be described by an expression of the form
\begin{center}
$
\begin{array}[t]{l}
h \scomp (\idn{k} \pcomp \cp{m-(k+l)} \pcomp \idn{l}) \scomp
(f \pcomp g) \scomp
(\idn{k'} \pcomp \eq{n-(k'+l')} \pcomp \idn{l'}) \scomp h'\;,
\end{array}
$
\end{center}
where
$f  \in \rel(S)(m-l,n-l')$ is a proper stream transformer,
$g  \in \rel(S)(m-k,\linebreak[2]n-k')$ is a direct connection, and
$h  \in \rel(S)(m,m)$ and
$h' \in \rel(S)(n,n)$ are bijective direct connections.
The constants $\cp{n} \in \rel(S)(n,n+n)$ and
$\eq{n} \in \rel(S)(n+n,n)$ used here are the ones defined below in
Definition~\ref{dfn-qrel}.
The restriction of $\rel(S)$ to quasiproper stream transformers is
denoted by $\qrel(S)$.
The further restriction of $\qrel(S)$ to functions is denoted by
$\qfn(S)$.
\edfn

With $\qfn(S)$ only deterministic synchronous dataflow networks can be 
modelled, whereas $\qrel(S)$ covers non-deterministic synchronous 
dataflow as well.
If $S$ is a set of streams of data, i.e.\
$S = \iseqof{(D \union \set{\tick})}$ for some set of data $D$, the
constants of BNA as defined on $\rel(S)$ in Section~\ref{bna-rel} are
quasiproper functions.
So the identity and transposition constants are in $\qfn(S)$ and
$\qrel(S)$.
In addition, both $\qfn(S)$ and $\qrel(S)$ are closed under the parallel
and sequential composition operations as defined on $\rel(S)$.
As mentioned before, the feedback operation as defined on $\rel(S)$
does not model feedback in synchronous dataflow networks properly.
A related problem is that $\qfn(S)$ is not closed under this feedback
operation.
All this means that only a more appropriate feedback operation and the
additional constants for synchronous dataflow have to be defined.

\bdfn (stream transformer model for synchronous dataflow)
\label{dfn-qrel}

\noindent
The parallel and sequential composition operations on $\qrel(S)$ are the
restrictions of the parallel and sequential composition operations on
$\rel(S)$ to $\qrel(S)$.
The identity and transposition constants in $\qrel(S)$ are the ones in
$\rel(S)$.

The feedback operation is redefined on $\qrel(S)$ as follows:
\begin{center}
\footnotesize
\begin{tabular}{lll}
\multicolumn{3}{l}{Notation} \\[-1.25ex]
\multicolumn{3}{l}{\rule{.99\textwidth}{.125mm}} \svsp \\
$f \feed{p}$ & $\in \qrel(S)(m,n)$ & for $f \in \qrel(S)(m+p,n+p)$ \\
\multicolumn{3}{l}{\rule{.99\textwidth}{.125mm}}
\end{tabular}
\svsp \\
$
\begin{array}{lll}
\multicolumn{3}{l}{\mbox{Definition}} \\[-1.25ex]
\multicolumn{3}{l}{\rule{.99\textwidth}{.125mm}} \svsp \\
f \feed{1} & = &
\begin{array}[t]{ll}
\set{\pair{x}{y} \where
     x \in S^m \And y \in S^n \And
     \exists z \in S \st \pair{x \cat z}{y \cat z} \in f}
& \mbox{if } (m+1,n+1) \notin dc(f) \\
(\idn{m} \pcomp \asour{1}) \scomp f \scomp (\idn{n} \pcomp \ssink{1})
& \mbox{otherwise}
\end{array}
\vsp \\
\multicolumn{3}{l}{
\mbox{for $p \neq 1$, $\feed{p}$ is defined by the equations occurring
as axioms R5--R6 of BNA}} \\
\multicolumn{3}{l}{\rule{.99\textwidth}{.125mm}}
\end{array}
$
\end{center}
The constants $\asour{n} \in \qrel(S)(0,n)$ and
$\ssink{n} \in \qrel(S)(n,0)$ used here are the ones defined right
away.

The additional constants for synchronous dataflow are defined on
$\qrel(S)$ as follows:
\begin{center}
\footnotesize
\begin{tabular}{lll}
\multicolumn{2}{l}{Notation} \\[-1.25ex]
\multicolumn{2}{l}{\rule{.99\textwidth}{.125mm}} \svsp \\
 $\cp{n}$     & $\in \qrel(S)(n,n+n)$           \\
 $\ssink{n}$  & $\in \qrel (S)(n,0)$            \\
 $\eq{n}$     & $\in \qrel(S)(n+n,n)$           \\
 $\asour{n}$  & $\in \qrel (S)(0,n)$            \\
\multicolumn{2}{l}{\rule{.99\textwidth}{.125mm}}
\end{tabular}
\svsp \\
$
\begin{array}{lll}
\multicolumn{3}{l}{\mbox{Definition}} \\[-1.25ex]
\multicolumn{3}{l}{\rule{.99\textwidth}{.125mm}} \svsp \\
\cp{n}     & = & \set{\pair{x}{x \cat x} \where x \in S^n} \\
\ssink{n}  & = & \set{\pair{x}{()}   \where x \in S^n} \\
\eq{n}     & = &
\set{\pair{(x_1,\ldots,x_n,y_1,\dots,y_n)}{(x_1 \eqt y_1,\dots,x_n \eqt y_n)}
     \where (x_1,\ldots,x_n), (y_1,\dots,y_n) \in S^n} \\
           &   &
\mbox{where  $(x \eqt y)(k) = x(k)$  if $x(k) = y(k)$ and
             $(x \eqt y)(k) = \tick$ otherwise} \\
\asour{n}  & = & \set{\pair{()}{(\tick^\infty,\ldots,\tick^\infty)}} \\
\multicolumn{3}{l}{\rule{.99\textwidth}{.125mm}}
\end{array}
$
\end{center}
\edfn

In Definition~\ref{dfn-rel}, the feedback operation was defined such
that, for each data transformer $f$, the feedback loop behaves as the
greatest fixpoint of $f$ relative to the input stream of $f \feed{}$.
In case of proper stream transformers, there is always a unique
fixpoint provided the transformer is a function or a continuous
relation (with respect to the prefixes of streams).
It means that the feedback loop is also the least fixpoint.
This is needed to model feedback in synchronous dataflow networks
properly; for otherwise it does not agree with the operational
understanding that it is iteratively feeding the network concerned with
data produced by it in the previous step.
The adaptation of the feedback operation given in
Definition~\ref{dfn-qrel} is needed to get a unique fixpoint in
case of quasiproper stream transformers as well.
It also guarantees that $\qfn(S)$ is closed under feedback.
Because $\cp{} \feed{}$ now produces a dummy stream, it equals the
dummy source.
For this reason, $\asour{}$ is used instead of $\ssour{}$ as constant
for synchronous dataflow.
Note that this stream transformer model does not have the global crash
property of the data transformer model from Section~\ref{bna-rel}: if a
component of a network fails to produce output on some tick of the
global clock, the effect is merely that the components connected to the
port(s) concerned will fail to produce output on some future tick.

\bthm
$(\qfn(S),\pcomp,\scomp,\feed{},\idn{},\tr{}{})$ is a model of BNA.
The constants $\cp{},\ssink{},\eq{},\asour{}$ satisfy the additional
axioms for synchronous dataflow networks (Table~\ref{tbl-na-s}).
\ethm
\bproof
For the first part, it is enough to prove R1--R4 and F1--F2.
According to~\cite{CS88,CS89}, it suffices to prove R1--R4 for $m = 1$,
and R4 additionally for $k = l = 1$ and $g = \tr{1}{1}$.
The proofs concerned are straightforward proofs by case distinction --
the cases depending on whether the ports relevant to the feedback loop
are directly connected or not.
The second part is a matter of tedious, but simple calculation.
\eproof

\subsection{Process algebra model for synchronous dataflow}
\label{na-s-sproc}
In this subsection, the specialization of the process algebra model of
BNA (Section~\ref{bna-proc}) for synchronous dataflow networks is given.
In this case, we will make use of \acpdrtt.
Recall that \acpdrtt\ is \acpdrt\ -- the discrete relative time
extension of ACP -- extended with abstraction based on branching
bisimulation.

In Section~\ref{bna-proc}, only a few assumptions about wires and atomic
cells were made.
Here it is first explained how these ingredients are actualized for
synchronous dataflow networks.
Because of the crucial role of the time slices determined by the ticks
of a global clock, discrete-time process algebra is used.
\bdfn (wires and atomic cells in synchronous dataflow networks)
\label{dfn-wires-cells-s}

\noindent
In the synchronous case, the identity constant, called the
{\em minimal stream delayer}, is the wire $\idn{1} = (1,1,\msd)$
where $\msd$ is defined by
$$
\msd =
\csl{\tau} \seqc (\asl{er}_1(x) \pref \csl{s}_1(x)) \seqc \delay(\msd)\;.
$$
The constants $\idn{n}$, for $n \neq 1$, and $\tr{m}{n}$ are defined by
the equations occurring as axioms B6 and B8--B9, respectively, of
Table~\ref{tbl-bna}.

In the synchronous case, the deterministic cell computing a function
$f:D^m \to D^n$, and having $\vec{a} = (a_1,\dots,a_n) \in D^n$ as its
initial output tuple, is the network $C_f(\vec{a}) = (m,n,P_f(\vec{a}))$
where $P_f$ is defined by
\begin{center}
$
\begin{array}[t]{l}
 P_f(\vec{a}) =
 \csl{\tau} \seqc
 (Out(\vec{a}) \cfm
  ((\asl{er}_1(x_1) \parc \dots \parc \asl{er}_m(x_m)) \pref
   \delay(P_f(f(x_1,\dots,x_m)))))\;, \vsp \\
 \mbox{where }
 Out(\vec{a}) =
 \csl{s}_1(a_1) \parc \dots \parc \csl{s}_n(a_n)\;.
\end{array}
$
\end{center}
The non-deterministic cell computing a (finitely branching) relation
$R \subseteq D^m \x D^n$, and having $A \subseteq D^n$ as its set of
possible initial output tuples, is the network $C_R(A) = (m,n,P_R(A))$
where $P_R$ is defined by
\begin{center}
$
\begin{array}[t]{l}
 P_R(A) =
 \csl{\tau} \seqc
 (Out(A) \cfm
  ((\asl{er}_1(x_1) \parc \dots \parc \asl{er}_m(x_m)) \pref
   \delay(P_R(R(x_1,\dots,x_m)))))\;, \vsp \\
 \mbox{where }
 Out(A) =
 \csl{\tau} \cond{A = \emptyset}
 \Altc{(a_1,\dots,a_n) \in A}
  (\csl{s}_1(a_1) \parc \dots \parc \csl{s}_1(a_n))\;.
\end{array}
$
\end{center}

The restriction of $\proc(D)$ to the processes that can be built under
this actualization is denoted by $\sproc(D)$.
\edfn

The definition of $\msd$ given above expresses the following.
The process $\msd$ waits until a datum is offered at its input port.
When a datum is available at the input port, $\msd$ delivers the datum
at its output port in the same time slice.
{From} the next time slice, it proceeds with repeating itself.

\noindent
The definition of $P_f$ expresses the following.
In the current time slice $P_f(\vec{a})$ produces the data
$a_1,\dots,a_n$ at the output ports $1,\dots,n$, respectively.
In parallel, $P_f(\vec{a})$ waits until one datum is offered at each of
the input ports $1,\dots,m$.
The waiting may last into subsequent time slices.
When data are available at all input ports, $P_f(\vec{a})$ proceeds with
repeating itself from the next time slice with a new output tuple, viz.\
the value of the function $f$ for the consumed input tuple.
The non-deterministic case ($P_R$) is similar.

For $\sproc(D)$, the operations and constants of BNA as defined on
$\proc(D)$ can be taken with $\msd$ as wire.
This means that only the additional constants for synchronous dataflow
have to be defined.

\bdfn (process algebra model for synchronous dataflow)
\label{dfn-sproc}

\noindent
The operations $\pcomp$, $\scomp$, $\feed{n}$ on $\sproc(D)$ are the
instances of the ones defined on $\proc(D)$ for $\msd$ as wire.
Analogously, the constants $\idn{n}$ and $\tr{m}{n}$ in $\sproc(D)$ are
the instances of the ones defined on $\proc(D)$ for $\msd$ as wire.

The additional constants in $\sproc(D)$ are defined as follows:
\begin{center}
\footnotesize
\begin{tabular}{ll}
\multicolumn{2}{l}{Notation} \\[-1.25ex]
\multicolumn{2}{l}{\rule{.99\textwidth}{.125mm}} \svsp \\
$\cp{1}$    & $\in \sproc(D)(1,2)$ \\
$\ssink{1}$ & $\in \sproc(D)(1,0)$ \\
$\eq{1}$    & $\in \sproc(D)(2,1)$ \\
$\asour{1}$ & $\in \sproc(D)(0,1)$ \\
\multicolumn{2}{l}{\rule{.99\textwidth}{.125mm}}
\end{tabular}
\svsp \\
$
\begin{array}{llll}
\multicolumn{4}{l}{\mbox{Definition}} \\[-1.25ex]
\multicolumn{4}{l}{\rule{.99\textwidth}{.125mm}} \svsp \\
\cp{1}    & = & (1,2,copy^1),   &
\multicolumn{1}{l}{\mbox{where }
 copy^1 =
  \csl{\tau} \seqc
  (\asl{er}_1(x) \pref (\csl{s}_1(x) \parc \csl{s}_2(x))) \seqc
  \delay(copy^1)} \\
\ssink{1} & = & (1,0,sink^1),   &
\multicolumn{1}{l}{\mbox{where }
 sink^1 =
  \csl{\tau} \seqc
  (\asl{er}_1(x) \pref \csl{\tau}) \seqc \delay(sink^1)} \\
\eq{1}    & = & (2,1,eq_1),     &
\multicolumn{1}{l}{\mbox{where }
 eq_1 =
  \csl{\tau} \seqc
  (\asl{er}_1(x_1) \pref P_2(x_1) \altc
   \asl{er}_2(x_2) \pref P_1(x_2)) \mbox{ and}} \\
\multicolumn{4}{r}{
 P_i(x) =
      \delay(eq_1) \altc
      \csl{{er}}_i(y) \pref (\csl{s}_1(x) \cond{x = y} \csl{\tau}) \seqc
      \delay(eq_1)
 \mbox{ for $i \in [2]$}} \\
\asour{1} & = & (0,1,source_1), &
\multicolumn{1}{l}{\mbox{where }
 source_1 = \csl{\tau} \seqc \asl{\dead}} \vsp \\
\multicolumn{4}{l}{
\mbox{for $n \neq 1$, these constants are defined by the equations
occurring as axioms A12--A19}} \\
\multicolumn{4}{l}{\mbox{in Table~\ref{tbl-na-s}}} \\
\multicolumn{4}{l}{\rule{.99\textwidth}{.125mm}}
\end{array}
$
\end{center}
\edfn
The equality test $\eq{1}$ does not necessarily perform one test per
time slice; it does so in order not to cause a time delay.
The definition of $eq_1$ expresses the following.
The process $eq_1$ waits until a datum is offered at one of its input
ports.
When a datum is available at one input port, it waits till the end of
the time slice concerned for a datum at the other port.
If this happens, it tests the equality of the data, delivers either in
case the test succeeds, and then proceeds with repeating itself from the
next time slice.
Otherwise, it skips the equality test and proceeds with repeating itself
from the next time slice.

The simpler equality test $\ol{\eq{}}_1 = (2,1,\ol{eq}_1)$, where
$$
\ol{eq}_1 =
\csl{\tau} \seqc
((\asl{er}_1(x) \parc \asl{er}_2(y)) \pref
 (\csl{s}_1(x) \cond{x = y} \csl{\tau}) \seqc \delay(\ol{eq}_1))\;,
$$
is not appropriate.
This equality test does not let data always pass through it with a
neglectible delay.
This means that it does not behave properly if the feedback operation
is applied; $\ol{\eq{}}_1\feed{1}$ is the process that deadlocks after
having read one datum -- it is a kind of dummy sink.
This failure to consume data does not fit in with the idea of permanent
flows of data which underlies synchronous dataflow.

\blem
\label{lem-msd}
The wire $\idn{1} = (1,1,\msd)$ gives an identity flow of data, i.e.\
for all $f = (m,n,P)$ in $\sproc(D)$,
$\idn{m} \scomp f = f = f \scomp \idn{n}$.
\elem
\bproof
It suffices to show that these equations hold for the atomic cells and
the constants.
The result then follows by induction on the construction of a network
in $\sproc(D)$.
$\idn{n} \scomp \idn{n} = \idn{n}$ and
$\tr{m}{n} \scomp \idn{n} = \tr{m}{n} = \idn{m} \scomp \tr{m}{n}$ follow
trivially from $\idn{1} \scomp \idn{1} = \idn{1}$.
For a proof of $\idn{1} \scomp \idn{1} = \idn{1}$, we refer
to~\cite{BB95c}.
So the asserted equations hold for $\idn{n}$ and $\tr{m}{n}$.
The proof for the remaining constants and the atomic cells is a
laborious piece of work in the same style.
\eproof
\bthm
\label{thm-sproc}
$(\sproc(D),\pcomp,\scomp,\feed{},\idn{},\tr{}{})$ is a model of BNA.
The constants $\cp{}$, $\ssink{}$, $\eq{}$, $\asour{}$ satisfy the 
additional axioms for synchronous dataflow networks 
(Table~\ref{tbl-na-s}).
\ethm
\bproof
A simple calculation shows that
$\idn{0} \pcomp f = f = f \pcomp \idn{0}$ for all $f \in \sproc(D)$.
The first part then follows immediately from Theorem~\ref{thm-proc} and
Lemma~\ref{lem-msd}.
The proof of the second part is a matter of tedious, but unproblematic
calculation in the style of~\cite{BB95c}.
\eproof

\bthm
\label{thm-sproc-complete}
The axioms in Table~\ref{tbl-na-s} are complete for closed terms.
\ethm
\bproof
For the proof of this theorem, we refer to~\cite{BS95a}.
\eproof

Queues that deliver data with a neglectible delay and never contain
more than one datum are an idealized concept; they do not occur in
practice.
More practical are wires that are interpreted as bounded queues.
It seems that bounded queues are most easily modelled as components
of asynchronous dataflow networks.

\section{Closing remarks}
\label{conclusions}
Concerning connections with earlier work on dataflow some additional
remarks are in order.

In~\cite{BWM94} a model for synchronous dataflow networks is presented.
Our Section~\ref{na-s-qrel} on a stream transformer model for
synchronous dataflow can be seen as a rephrasing of this work.
We consider the stream transformer model described in
Section~\ref{na-s-qrel} to be more denotational and the process algebra
model described in Section~\ref{na-s-sproc} to be more operational.

The model presented in~\cite{BWM94} is essentially a BNA model, although
it has some slightly different operations and constants.
For example, it has ``left-feedback'' ($\iter$) instead of
``right-feedback'' (see also the table below) and ``input sharing''
($\,^{\wedge}\,$) instead of the constants $\cp{}$ and $\tr{}{}$.
However, the constants and operations of BNA are definable in terms of
the ones of this model and vice versa.
%
The setting of~\cite{BWM94} may be obtained from our general network
algebra setting by taking BNA with the following parameters:
(1) the set of data $D$ is $\Nat$;
(2) the atomic cells are ``successor'' and ``conditional'';
(3) the additional constants for branching connections are
    $\cp{}$, $\ssink{}$ and $\eq{}$.
Kahn's history model~\cite{Kah74} is also essentially a BNA model (with
$\cp{}$, $\ssink{}$ and $\asour{}$ as additional constants) and so are
Broy's oracle based models~\cite{Bro88}.
SCAs~\cite{TT91} require for each internal stream in a network an
initial value.
We have taken that viewpoint as well.

Both the left- and right-feedback can be used.
The left-feedback can be defined in terms of the right-feedback as
follows:
$$
\feed{p} f =
 (\tr{p}{m} \circ f \circ \tr{p}{n}) \feed{p},\hsp f : p+m \to p+n\;.
$$
Other proposed feedback-like operators can be defined in terms of left-
or right-feedback:
\begin{center}
\footnotesize
\begin{tabular}{l@{\,\,\,\,}c@{\,\,\,\,}l@{\,\,\,\,}l@{}}
Name & Symbol & Network algebra definition & Ref. \\[-1.25ex] 
\multicolumn{4}{l}{\rule{.99\textwidth}{.125mm}} \svsp \\
feedback & $^*$ 
 & $f^*       = \feed{1} f,                  \hsp f : 1+m \to 1+n$
 & \cite{BWM94} \\ 
feedback & $\mu$ 
 & $\mu f     = (f \scomp \rmfii{m}) \feed{m}, \hsp f : n+m \to m$
 & \cite{Bro93} \\ 
(unary) star & $^*$ 
 & $f^*       = \rmfii{1} \scomp
                (\idn{1} \pcomp
                 (\idfii{1} \scomp f \scomp \rmfii{1}) \feed{1}) \scomp
                \idfii{1},                     \hsp f:1 \to 1$
 & \cite{CEW58}  \\ 
iteration & $\dagger$ 
 & $f^\dagger = \feed{m} (\idfii{m} \scomp f), \hsp f : m \to m+n$
 & \cite{Elg75}  \\ 
(binary) star & $^*$ 
 & $f^*g      = \rmfii{1} \scomp
                (\idn{1} \pcomp
                 \feed{1} (\idfii{1} \scomp f \scomp \rmfii{1})) \scomp
                \idfii{1} \scomp g,            \hsp f,g : 1 \to 1$
 & \cite{Kle56}  \\ 
\multicolumn{4}{l}{\rule{.99\textwidth}{.125mm}} 
\end{tabular}
\end{center}

\subsubsection*{Acknowledgements}
The understanding on dataflow computation of the third author was much
clarified by discussions with M.~Broy and K.~St{\o}len.
The first author acknowledges discussions with J.V.~Tucker on SCAs.

\bibliographystyle{splncs03}
\bibliography{NA}

\begin{thebibliography}{10}
\providecommand{\url}[1]{\texttt{#1}}
\providecommand{\urlprefix}{URL }

\bibitem{BB94a}
Baeten, J.C.M., Bergstra, J.A.: On sequential composition, action prefixes and
  process prefix. Formal Aspects of Computing  6,  \mbox{250--268} (1994)

\bibitem{BB95c}
Baeten, J.C.M., Bergstra, J.A.: Some simple calculations in relative time
  process algebra. In: Aarts, E.H.L. et al. (eds.) Simplex Sigillum Veri: A
  Liber Amicorum for Prof.\ F.E.J. Kruseman Aretz. Department of Computer
  Science, Eindhoven University of Technology (1995)

\bibitem{BM02a}
Baeten, J.C.M., Middelburg, C.A.: Process Algebra with Timing. Monographs in
  Theoretical Computer Science, An EATCS Series, Springer-Verlag (2002)

\bibitem{BW90}
Baeten, J.C.M., Weijland, W.P.: Process Algebra. Cambridge Tracts in
  Theoretical Computer Science~18, Cambridge University Press (1990)

\bibitem{BWM94}
Barendregt, H., Wupper, H., Mulder, H.: Computable processes. Tech. Rep.
  CSI-R9405, Computing Science Institute, Catholic University of Nijmegen
  (1994)

\bibitem{BK84b}
Bergstra, J.A., Klop, J.W.: Process algebra for synchronous communication.
  Information and Control  60,  \mbox{109--137} (1984)

\bibitem{BMS95a}
Bergstra, J.A., Middelburg, C.A., {\c{S}tef\u{a}nescu}, G.: Network algebra for
  synchronous and asynchronous dataflow. Report P9508, Programming Research
  Group, University of Amsterdam (1995)

\bibitem{BMS97a}
Bergstra, J.A., Middelburg, C.A., {\c{S}}tef{\u{a}}nescu, G.: Network algebra
  for asynchronous dataflow. International Journal of Computer Mathematics  65,
   57--88 (1997)

\bibitem{BS95a}
Bergstra, J.A., {\c{S}}tef{\u{a}}nescu, G.: Network algebra with demonic
  relation operators. Report P9509, Programming Research Group, University of
  Amsterdam (1995)

\bibitem{Boh84}
B{\"{o}}hm, A.P.W.: Dataflow Computation. CWI Tracts~6, Centre for Mathematics
  and Computer Science, Amsterdam (1984)

\bibitem{BA81}
Brock, J.D., Ackermann, W.B.: Scenarios: A model of non-determinate
  computation. In: Diaz, J., Ramos, I. (eds.) Formalisation of Programming
  Concepts. pp. 252--259. LNCS~107, Springer-Verlag (1981)

\bibitem{Bro88}
Broy, M.: Nondeterministic dataflow programs: How to avoid the merge anomaly.
  Science of Computer Programming  10,  65--85 (1988)

\bibitem{Bro93}
Broy, M.: Functional specification of time sensitive communicating systems. ACM
  Transactions on Software Engineering and Methodology  2,  1--46 (1993)

\bibitem{CS88}
C{\u{a}}z{\u{a}}nescu, V.E., {\c{S}}tef{\u{a}}nescu, G.: A formal
  representation of flowchart schemes~{I}. Analele Universit{\u{a}}tii
  Bucuresti, Matematic{\u{a}} - Informatic{\u{a}}  37,  33--51 (1988)

\bibitem{CS89}
C{\u{a}}z{\u{a}}nescu, V.E., {\c{S}}tef{\u{a}}nescu, G.: A formal
  representation of flowchart schemes~{II}. Studii si Cercet{\u{a}}ri
  Metematice  41,  151--167 (1989)

\bibitem{CS90}
C{\u{a}}z{\u{a}}nescu, V.E., {\c{S}}tef{\u{a}}nescu, G.: Towards a new
  algebraic foundation of flowchart scheme theory. Fundamenta Informaticae  13,
   171--210 (1990)

\bibitem{CEW58}
Copy, I.M., Elgot, C.C., Wright, J.B.: Realization of events by logical nets.
  Journal of the ACM  5,  181--196 (1958)

\bibitem{Elg75}
Elgot, C.C.: Monadic computation and iterative algebraic theories. In: Rose,
  H.E., Sheperdson, J.C. (eds.) Logic Colloquium~'73. pp. 175--230. Studies in
  Logic and the Foundations of Mathematics, Volume~80, North-Holland (1975)

\bibitem{GW96}
van Glabbeek, R.J., Weijland, W.P.: Branching time and abstraction in
  bisimulation semantics. Journal of the ACM  43(3),  555--600 (1996)

\bibitem{Jon89}
Jonsson, B.: A fully abstract trace model for dataflow and asynchronous
  networks. Distributed Computing  7,  197--212 (1994)

\bibitem{Kah74}
Kahn, G.: The semantics of a simple language for parallel processing. In:
  Rosenfeld, J.L. (ed.) Information Processing~'74. pp. 471--475 (1974)

\bibitem{Kle56}
Kleene, S.C.: Representation of events in nerve nets and finite automata. In:
  Shannon, C.E., McCarthy, J. (eds.) Automata Studies. pp. 3--41. Annals of
  Mathematical Studies, Volume~34, Princeton University Press (1956)

\bibitem{Kok87}
Kok, J.: A fully abstract semantics for data flow nets. In: de~Bakker, J.W.,
  Nijman, A.J., Treleaven, P.C. (eds.) PARLE~'87. pp. 351--368. LNCS~259,
  Springer-Verlag (1987)

\bibitem{Rus89}
Russell, J.: Full abstraction for nondeterministic dataflow networks. In:
  FoCS~'89. IEEE Computer Science Press (1989)

\bibitem{Ste87b}
{\c{S}}tef{\u{a}}nescu, G.: On flowchart theories: Part~{II}. {T}he
  nondeterministic case. Theoretical Computer Science  52,  307--340 (1987)

\bibitem{Ste86}
{\c{S}}tef{\u{a}}nescu, G.: Feedback theories (a calculus for isomorphism
  classes of flowchart schemes). Revue Roumaine de Mathematiques Pures et
  Applique  35,  73--79 (1990)

\bibitem{Ste94}
{\c{S}}tef{\u{a}}nescu, G.: Algebra of flownomials. {P}art~{1}: {B}inary
  flownomials, basic theory. Report TUM I9437, Department of Computer Science,
  Technical University Munich (1994)

\bibitem{TT91}
Thompson, B.C., Tucker, J.V.: Algebraic specification of synchronous concurrent
  algorithms and architecture. Tech. Rep. 10-91, Department of Mathematics and
  Computer Science, University College of Swansea (1991)

\end{thebibliography}

\end{document}